\documentclass[%
 reprint,
 amsmath,amssymb,
 aps,prl
]{revtex4-1}

\usepackage{mathtools}
\usepackage{multirow}
\usepackage{color}
\usepackage{epsf}
\usepackage{epsfig}
\usepackage{ulem}
\usepackage{graphicx}
\usepackage{dcolumn}
\usepackage{bm}

\begin{document}

\title{Electronic, structural, and magnetic properties of LaMnO$_3$ phase transition at high temperature}


\author{Pablo Rivero$^1$, Vincent Meunier$^2$, and William Shelton$^{1}$}

\email{jprivero@lsu.edu}

\affiliation{1. Center for Computation and Technology, Louisiana State University, Baton Rouge, Louisiana 70803, USA \\
2. Department of Physics, Applied Physics, and Astronomy, Rensselaer Polytechnic Institute, Troy, NY 12180, USA }

\date{\today}

\begin{abstract}

\noindent{}We develop a procedure to determine the portion of exact Hartree-Fock exchange interaction contained in a hybrid density functional to treat the range of electronic correlation governing the physics of a system as a function of a thermodynamical parameter. This includes systems that depend on physical parameters accessible to experiment (i.e., temperature, pressure, composition, etc.) or those composed of two or more materials such as heterostructures and interfaces. This approach is applied to LaMnO$_3$ where for the first time we are able to simulate the high temperature insulator-to-metal transition (IMT) and observe a half-metallic orbital disorder ferromagnetic state using density functional theory. In particular, we show that the softening of the \textit{Q$_2$} Jahn-Teller mode plays a central role in driving the IMT. These findings are likely to motivate the investigation of heterostructures and bulk materials that contain a range of electronic correlation in similar material systems. 

\begin{description}
\item[PACS numbers]
71.30.+h, 31.15.A-, 71.20.Be, 71.70.Ej

\end{description}
\end{abstract}

\pacs{Valid PACS appear here}
\maketitle


\section{I. INTRODUCTION}

LaMnO$_3$ perovskite-type manganites have been subjected to intense scrutiny due to the existence of highly desired properties such as the observed colossal magnetoresistance (CMR) by either introducing impurities \cite{CMR1,CMR2,LSMO} or applying pressure \cite{pressure}. This effect along with the occurrence of fully spin-polarized conduction bands \cite{halfmetallic} makes these materials ideal candidates for the development of spintronic applications \cite{spintronic1,spintronic2}. LaMnO$_3$ exhibits a range of electronic correlation (i.e., different strength of the electronic interactions in the system) as a function of temperature. At low temperature the system is a Mott-Hubbard A-type antiferromagnetic (A-AFM) insulator that crystallizes in the Pbnm orthorhombic structure \cite{exp0}. However, at temperatures above T$_{JT}$ = 750 K the system forms a \textit{pseudo-cubic} ferromagnetic (FM) metal \cite{FM2,metal1,metalferro} where $a$ and $b$ lattice parameters are nearly equal \cite{carvajal}. The transition to this high temperature phase involves a significant reduction of the Jahn-Teller (JT) distortions that leads to small decreases of the MnO$_6$ octahedra tilts and rotations \cite{carvajal} and insulator-to-metal transition (IMT). To fully appreciate the mechanism that stabilizes these properties, a complete understanding of the stoichiometric LaMnO$_3$ compound is needed.  

An accurate and consistent density functional theory (DFT) description of the underlying physics requires a procedure that provides a single functional that can be applied to systems with a range of electronic correlation. This functional must be able to describe the energetics and electronic structure associated with the strength of the electronic correlation for both the IMT and the magnetic transition observed at high T. For instance, the development of a DFT-based approach with an appropriate amount of non-local Hartree-Fock exchange (HFX) could allow for the direct calculation of the energetics and electronic structure of both high and low temperature phases, including those associated with JT distortions. Hybrid functionals include a HFX treatment of the Kohn-Sham orbitals that makes them attractive candidates to meet the above requirements. Using the same hybrid functional to simulate the electronic and magnetic states can provide a consistent way for comparing the electronic properties thus, allowing for a consistent interpretation of the mechanism associated with the IMT, magnetic and structural phase changes.

In this report we describe a procedure to determine the necessary amount of HFX mixing contained in a particular hybrid DFT functional to capture both the LaMnO$_3$ A-AFM insulating ground-state and the FM metallic behavior at high T. This type of approach provides a path for studying systems with a range of electronic correlation as a function of experimentally controlled parameters. This includes bulk systems such as ruthenates and low-dimensional systems such as SrTiO$_3$/La$_{1-x}$Sr$_x$MnO$_3$ interfaces, and surfaces.

At low temperature, the MnO$_6$ octahedra of LaMnO$_3$ are tilted, rotated and distorted by JT instabilities characterized by two short ($s$ = 1.903 \r{A}), two long ($l$ = 2.184 \r{A}), and two medium Mn-O bond distances ($m$ = 1.957 \r{A}) \cite{exp0}. These instabilities lead to a lifting of the degeneracy of the twofold degenerate e$_g$ orbital (singly occupied) that stabilizes the system into an orbital ordered type insulating state \cite{ord1,ord2}. The magnetic phase is A-AFM, where Mn atoms are coupled ferromagnetically in the $ab$-plane and antiferromagnetically along the $c$-direction (see Fig. 1). There are two symmetry-distinct types of oxygen atoms in this structure. The first type, $O1$, lies in the $ab$-plane and is characterized by one short and one long Mn-O bonds. The second type, $O2$, is oriented along the $c$-direction and provides two medium size Mn-O bonds.

The magnetic ordering of this phase is consistent with the Goodenough-Kanamori rules \cite{good,kanamori2}. These rules are based on the symmetry relations and electron occupancy of the overlapping atomic orbitals. They predict that the FM ordering observed in the $ab$-plane are due to the superexchange associated with the Mn-$O1$ bonds and the e$_g$ partial occupation, while in the $c$-direction the isotropic Mn-$O2$ bonds lead to an AFM superexchange interaction.

\begin{figure}[t]
\includegraphics[width=0.48\textwidth]{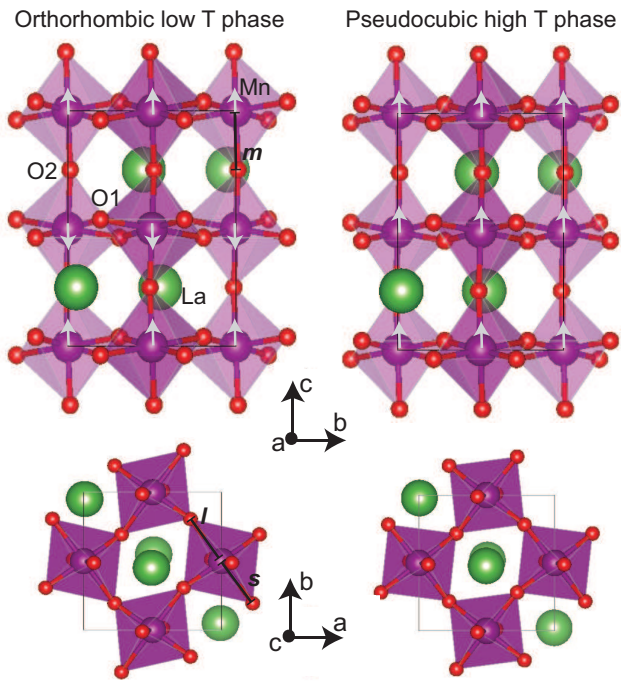}
\caption{Side and top views of the orthorhombic low T phase (left) and the \textit{pseudo-cubic} high T phase (right) of LaMnO$_3$ structure. Long (l), medium (m), and short (s) Mn-O bonds as well as the spin ordering are shown.}
\label{fig1}
\end{figure}

Upon heating the system above the N\'eel temperature (140K) \cite{Neel}, LaMnO$_3$ transforms from the A-AFM state into a paramagnetic state characterized by time-fluctuating short-range spin ordering between Mn atoms \cite{paramagnetic} consistent with the Curie-Weiss law. At T$_{JT}$  = 750 K a first order structural phase transition takes place with both an IMT \cite{metal1,metalferro} and an orbital order-disorder transition \cite{ESR}. These transitions are driven by the reduction of the cooperative JT distortions \cite{ord1,carvajal} which result in a crystal structure similar to the Pbnm crystallographic symmetry but with nearly equal $a$ and $b$ lattice constants (referred as \textit{pseudo-cubic}). Moreover, the Weiss temperature constant becomes positive upon crossing T$_{JT}$, thereby indicating the existence of FM interactions in LaMnO$_3$ \cite{metalferro,FM2}.

There has been considerable efforts to understand the IMT in LaMnO$_3$. It is known that different types of IMT can be driven by different experimentally controlled parameters: pressure, temperature, or doping. Experimental studies carried out using low-temperature Raman spectroscopy \cite{raman} and first-principles calculations \cite{franchini2} demonstrated that high hydrostatic pressure ($\sim$ 32GPa) reduces the JT distortions, which govern the IMT in LaMnO$_3$. Doping LaMnO$_3$ with divalent atoms (Sr or Ca) reduces the electron filling in the $d$-band and produces Mn$^{+4}$ ions. This eventually results in a new type of IMT \cite{LSMO}. In fact, for the limit of complete substitution of La with Sr the e$_g$ orbital is no longer occupied and thus, it is not subjected to JT distortions. Therefore, any variation of these parameters (T, P, or doping) can induce dramatic changes in the strong interplay of electron, spin, lattice, and orbital degrees of freedom. 

The common effect of these interactions is illustrated by the balance between the e$_g$ orbital degeneracy and the lattice strain energy compensation that results from the reduction of the JT distortions. This effect is the leading cause for the structural, magnetic, and IMT transitions. However, current understanding of the electronic origins of the IMT and the associated high-temperature phase of LaMnO$_3$ remains incomplete. The main difficulty lies in being able to simulate a system containing a range of electronic correlation as a function of experimentally controlled parameters within a single first-principles framework. One proper way to achieve a satisfactory description would be by the use of a single DFT hybrid based functional rather than a collection of \textit{ad hoc} parameters that can only account for physical behavior under a given set of external parameters.

Here, we demonstrate the existence of a single hybrid DFT functional with a particular HFX mixing that accounts for the detailed electronic origins of the structural, magnetic, and IMT in LaMnO$_3$. This approach makes use of local atom-centered basis that provides a natural screening of the HFX potential within a few neighboring shells of atoms. This important feature results in high computational performance that makes it possible to simulate large systems using hybrid DFT functionals. The numerical efficiency offers a path towards the simulation of low-dimensional systems such as LaMnO$_3$ based heterostructures where electronic correlation is known to play an important role on the observed interfacial properties \cite{okamoto}.

\section{II. METHODS AND PREVIOUS WORKS}

Simulations using DFT within the semi-local GGA have been performed previously. However, they predicted a metallic FM ground-state instead of the experimentally observed insulating A-AFM phase \cite{franchini,pbebad2,mellan}.  This failure has been attributed to the self-interaction problem plaguing DFT.

To accurately describe the properties of LaMnO$_3$ phases requires methods beyond DFT, such as the GGA+U (or LDA+U) approach. This approximation is based on a Hubbard-type Hamiltonian approach where an on-site constant effective potential is introduced to account for the intra-atomic electron Coulomb repulsive potential (U) and intra-orbital exchange potential (J) \cite{dudarev,lich}. The works of Hashimoto $et$ $al.$ \cite{pbebad2} and Mellan $et$ $al.$ \cite{mellan} confirmed an improvement of the description of the LaMnO$_3$ properties by means of an empirical fit of the U and J parameters. However, this method is truly a local approach for describing strongly correlated systems.


Hybrid functionals offer a good alternative to DFT and DFT+U approximations. They allow for different amounts of HFX to be mixed with density functional approximations (DFA) (either LDA or GGA) exchange potentials. In general, the hybrid DFT exchange-correlation potential can be written as

\begin{equation}
V^{Hyb}_{XC}=\alpha(V^{HF}_{X})+(1-\alpha)(V^{DFA}_{X})+V^{DFA}_{C} 
\end{equation}

\noindent{}where \textit{V$^{DFA}_{X}$} and \textit{V$^{HF}_{X}$} are the DFA and HF exchange potentials respectively, $\alpha$ accounts for the percentage of HFX potential that is mixed with the DFT exchange potential, and \textit{V$^{DFA}_{C}$} is the DFA correlation potential. Unlike GGA+U (LDA+U), hybrid functionals account for both the electronic localization acting on all states of the system in an orbital dependent fashion and the non-local behavior that is not treated within the DFT+U on-site approach.

More elaborate hybrid functionals, called range-separated functionals, have provided excellent performance for describing relevant properties of a variety of chemical systems. These functionals make use of a modified form of the Coulomb operator separated into long-, medium-, and short-ranges where different percentages of \textit{V$^{HF}_{X}$} and \textit{V$^{DFA}_{X}$} apply. The separation, which is achieved via combination of error (erf) and complementary error (erfc) functions, is written as

\begin{equation}
\begin{multlined}
\frac{1}{r} = \frac{{\rm erfc}(\omega_{SR} r)}{r}+\frac{{\rm erf}(\omega_{LR} r)}{r} +\\
\left ( \frac{{\rm erfc}(\omega_{SR} r)}{r}+\frac{{\rm erfc}(\omega_{LR} r)}{r} \right)
\end{multlined}
\end{equation}

\noindent{}where $\omega$$_{SR}$ and $\omega$$_{LR}$ are parameters that define the length scale of separation for the three different regions. The first two terms represent the short- and long-ranges while the third term represents the middle-range.

There are various types of range-separated hybrid functionals, which depend on the number of regions that are used in splitting the Coulomb operator. For example, the Lc-$\omega$PBE \cite{lcwpbe} is the long-range separated functional, which uses 100$\%$ PBE (GGA) exchange for the short-range component and 100$\%$ HFX in the long-range part. On the other hand HSE \cite{hse06} is a short-range functional since it only contains 25$\%$ HFX locally, similar to PBE0 ($\alpha$ = 0.25), while it uses 100$\%$ PBE functional at long-range. Even though both HSE and Lc-$\omega$PBE perform well for atomic energies, heats of formation \cite{heats}, and magnetic properties \cite{magn1,magn2}, Lc-$\omega$PBE performs badly for bandgaps and HSE poorly for reaction barriers.

These shortcomings have lead to the development of a triple range-separated functional known as HISS \cite{hiss}. This multi-range separation functional uses 100$\%$ PBE in the short- and long-range regimes and 60$\%$ HFX in the middle-range. Over a couple of nearest-neighbor distances, HISS has more non-local HFX than HSE but the fraction of HFX in HISS decays more rapidly. Since the middle-range separation contains some overlap with both the short- and long-range components and because it contains considerable amount of HFX mixing, this functional usually predicts larger bandgaps and AFM-FM energy differences than HSE but smaller than Lc-$\omega$PBE.

Plane-wave based hybrid DFT calculations showed that a local mixing of 15$\%$ HFX in the HSE functional reproduces the properties of LaMnO$_3$ ground-state \cite{franchini}. In addition, this functional also reproduces the experimental IMT observed at high hydrostatic pressure on LaMnO$_3$ \cite{franchini2} and the doping induced IMT on BaBiO$_3$ \cite{BaBiO3}, supporting the idea of using a unique hybrid DFT functional to study phase transitions. However, the use of plane-wave basis sets is a computationally demanding approach since plane-waves are global basis sets in contrast to local atomic-centered basis sets that provide a natural screening of the HFX potential within a few neighboring shells of atoms. Thus, the use of plane-waves strongly limits the system size due to its poor scaling. Note that SIC-LDA \cite{lmno-sic} and GW \cite{GW,GW2} approaches have also been used to study the low temperature phase of LaMnO$_3$. However these calculations were not carried out for full geometry relaxation and therefore, it is not clear if the results correspond to the true ground-state of the system.

\section{III. COMPUTATIONAL DETAILS}

All calculations were performed with the replicated-data version of the CRYSTAL14 computational package \cite{CRYSTAL14,CRYSTAL142}. CRYSTAL14 is a first-principles electronic structure software that employs atom-centered Gaussian-type orbital (GTO) basis sets. GTOs are used to build Bloch functions to expand the one-electron crystalline orbitals. Interestingly, GTOs present a number of advantages including the use of local basis sets that contain minimal overlap with neighboring orbitals. This feature provides a significant improvement in computational scaling with increasing system size that allows for simulation of large-scale systems using DFT hybrid functionals. This high computational efficiency is particularly important even for moderate size systems containing tens of atoms. Indeed, while the HFX part of the simulation scales as $\textit{N$^4$}$ where $N$ is the number of atoms making up the system, the scaling is reduced to $\textit{N$^2$-N$^3$}$ for larger systems thanks to the built-in screening.

All-electron GTO basis sets for each atom comprising the LaMnO$_3$ system were chosen as follows: for Mn the 6-411d41G atomic basis set generated by Towler $et$ $al.$ \cite{Towler} was employed; for O the 8-411d basis set constructed by Cor\`a \cite{Cora} was used; for La, the 61111sp-631d basis set with 1s, 6sp, 3d contractions was obtained directly from the CRYSTAL14 website \cite{webpage}. We added two 4f electron shells to the La basis set to better account for the La-O and Mn-O covalency induced by the octahedral rotations found in this system  \cite{cov1,cov2}.

An 8 x 8 x 8 Monkhorst-Pack mesh \cite{monkhorst} that corresponds to 170 k-points in the irreducible Brillouin zone has been adopted.  We set the thresholds controlling the accuracy in the evaluation of Coulomb and exchange integrals to 10$^{-7}$ (ITOL1, ITOL2, ITOL3, and ITOL4, using notations from Ref. \onlinecite{CRYSTAL142}) and 10$^{-14}$ (ITOL5). The threshold on the SCF energy was set to 10$^{-6}$ au. Two different types of geometry optimizations have been performed depending on the LaMnO$_3$ phase with the same convergence criterion on gradient components and nuclear displacements set to 0.0003 Hartree/Bohr and 0.0012 Bohr respectively. For the low temperature orthorhombic phase a full relaxation scheme that optimizes both cell parameters and atomic coordinates was employed. In contrast, for the high temperature \textit{pseudo-cubic} phase a constrained full geometry relaxation was carried out by imposing $a$ = $b$ and $c$ = $a$ $\sqrt{2}$, which leads to a single parameter optimization. As will be shown, this approach produces an error inferior to 0.1$\%$ in the cell parameters compared to experiment.

The hybrid functionals studied here include B3LYP \cite{b3lyp}, B3PW \cite{pwgga}, the range separated hybrids: HSE \cite{hse06}, HISS \cite{hiss}, Lc-$\omega$PBE \cite{lcwpbe} and the M06 meta-GGA \cite{m06}. We also carefully examined the effect of using different amounts of mixing of the PBE \cite{pbe} exchange potential V$^{PBE}_{X}$ with Hartree-Fock exchange V$^{HF}_{X}$ potential (PBE-HFX). The PBE-HFX mixing is expressed as in eq. 1, where $\alpha$ = 0.05, 0.10, 0.15, 0.20 and 0.25 ($\alpha$  = 0.25 represents the PBE0 \cite{pbe0} hybrid functional). For convenience the PBE-HFX functionals will be written as PBE-5, PBE-10, PBE-15, PBE-20 and PBE-25 (PBE0) to represent the percentage of HFX mixing contained in them. Similar notation will be used for HSE (HSE-HFX).

\section{IV. RESULTS AND DISCUSSION}

\subsection{A. Performance of hybrid functionals for low T experimental structure}

We will first study the performance of different hybrid functionals for the description of the experimental ground-state structure. Then, the electronic and magnetic properties will be compared with the relaxed structure properties to evaluate how are they affected. This assessment will provide important information for down selecting candidate functionals to be used subsequently on the high temperature FM metallic phase. 

When determining the ground-state magnetic structure we considered all possible antiferromagnetic orderings (A-AFM, C-AFM, and G-AFM) and A-type was found to have the lowest energy for all DFT functionals. Therefore, AFM and A-AFM will be used interchangeably. Table I summarizes all numerical results obtained in our initial screening procedure applied to the experimental ground-state orthorhombic structure. It includes the total energy difference between AFM and FM phases (relative to the FM state), the bandgap, and the magnetic moments calculated for a variety of hybrid functionals as well as for the GGA (PBE) functional. We also include a number of results reported in the literature, namely plane-wave basis set results obtained via PBE, PBE+U \cite{mellan} and HSE \cite{franchini}.

These results establish that the A-AFM state is captured by PBE and all hybrid functionals with the notable exception of M06. Although PBE correctly describes the AFM state, it significantly underestimates the bandgap (0.19 eV). PBE+U approximation does produce an AFM insulating state for the optimal values of U = 8 eV and J = 2 eV with a significantly improved bandgap of 1.6 eV and a rather small 2 meV AFM-FM energy difference.

\begin{table}[]
\centering
\caption{$\Delta$E = E$_{AFM}$-E$_{FM}$ (meV), Mn magnetic moments ($\mu$ $_B$), and bandgap (eV) results calculated for the low T experimental crystal structure via different DFT approaches. }
\label{my-label1}
\scalebox{1.0}{
\begin{tabular}{lccc}
\\
\hline
    & $\Delta$E &    m  & Gap  \\ \hline 
PBE                  & -94 & 3.62   & 0.19     \\
PBE-5               & -81 & 3.71   & 0.62     \\
PBE-10              & -54 & 3.76   & 1.16     \\
PBE-15              & -39 & 3.80   & 1.66     \\
PBE-20              & -26 & 3.83   & 2.23     \\
PBE-25              & -20 & 3.84   & 2.79     \\
B3LYP               & -38 & 3.82   & 2.29     \\
B3PW                & -21 & 3.83   & 2.20     \\
HSE-25              & -11 & 3.85   & 2.17     \\
Lc-$\omega$PBE      & -77 & 3.75   & 6.88     \\
HISS                & -33 & 3.85   & 3.10     \\
M06                 & 4.0  & 3.87  & 1.17        \\ \hline \hline
PBE$^{a}$           & -63  & 3.50  & 0.23     \\
HSE-10$^{a}$        & -44  & 3.62  & 1.10     \\
HSE-15$^{a}$        & -28  & 3.67  & 1.52     \\
HSE-25$^{a}$        & -11  & 3.73  & 2.40     \\
PBE+U(U$_{eff}$=2)$^{b}$      &  1   & -     & 0.6      \\
PBE+U(U$_{eff}$=8)$^{b}$      &  17  & -     & 1.4      \\
PBE+U(U=8,J=2)$^{b}$ & -2   & -     & 1.6      \\ \hline
Exp.                 &  \textless 0   & 3.72$^{c}$,3.87$^{d}$  & 1.1$^{e}$,1.7$^{d}$,2.0$^{f}$ \\ \hline
\\
$^{a}$ \cite{franchini}
$^{b}$ \cite{mellan}
$^{c}$ \cite{exp1} \\
$^{d}$ \cite{exp2}
$^{e}$ \cite{exp3}
$^{f}$ \cite{exp4}

\end{tabular}}
\end{table}

The origins of the difference between the results obtained by PBE-25 and HSE-25 hybrid functionals lie in the amount of HFX that each functional contains for the long-range exchange interaction. PBE-25 has 25$\%$ mixture of HFX for any range while HSE-25 only contains the same mixture locally and 100$\%$ PBE exchange at long-range. Therefore PBE-25 predicts larger bandgaps and larger AFM-FM energy differences while HSE has a screened tail that favors a metallic behavior. Since both magnetic state and bandgaps are affected by structural changes, the quality and robustness of these functionals must be assessed for full geometry relaxation.

\subsection{B. Performance of hybrid functionals for full geometry relaxation}

The second step of our functional search will make it possible to determine the effective capability of the hybrid functionals and the optimal HFX mixing to obtain accurate results of the low temperature AFM insulating state. This determination will be performed by comparing the properties of fully relaxed structures directly to experiment. 

\begin{figure}[t]
\includegraphics[width=0.45\textwidth]{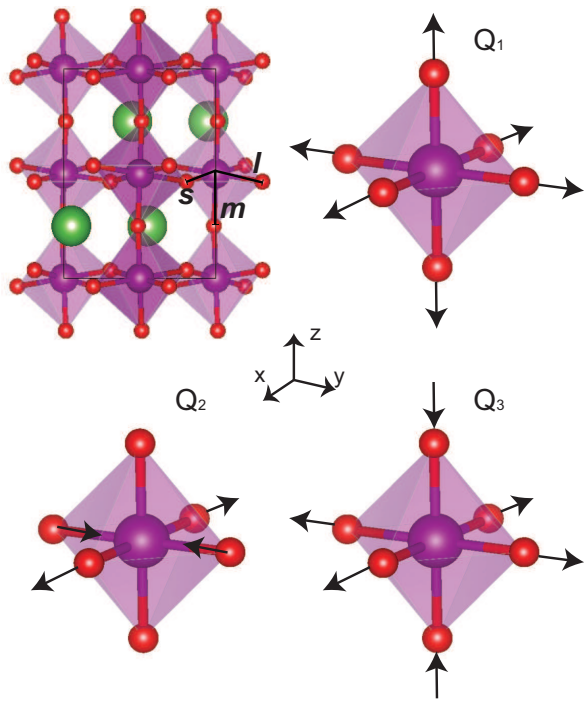}
\caption{Jahn Teller \textit{Q$_1$}, \textit{Q$_2$}, and \textit{Q$_3$} distortion modes in LaMnO$_3$ orthorhombic structure. The arrows indicate the displacements generated by the movement of the oxygen atoms with respect to the regular octahedron. The top left panel displays the $\it{l}$, $\it{m}$, and $\it{s}$ distorted Mn-O bonds.}
\label{fig2}
\end{figure}

In LaMnO$_3$, the MnO$_6$ octahedra contain three symmetrical vibrational modes (see Fig. 2) named \textit{Q$_1$}, \textit{Q$_2$}, and \textit{Q$_3$} \cite{kanamori,vanvleck} that correspond to the three JT distortions. \textit{Q$_1$} represents a breathing mode while \textit{Q$_2$} and \textit{Q$_3$} are lattice distortions, which produce different Mn-O bond distances (long, medium, and short) and therefore are the principal modes associated with the JT distortions. These two modes can be expressed as

\begin{equation}
Q_2 = \frac{2(l-s)}{\sqrt{2}}
\end{equation}

\begin{equation}
Q_3 = \frac{2(2m-l-s)}{\sqrt{6}} 
\end{equation}

\noindent{}where $\it{l}$, $\it{m}$, and $\it{s}$ refer to the long, medium, and short Mn-O distances, respectively.

JT distortions induce a local change in symmetry that leads to a lifting of the degeneracy of the doubly degenerate e$_g$ bands, and the opening of a bandgap. This results in lowering the total energy of the system. MnO$_6$ octahedra are connected and cooperative distortions are expected, generating alternated long and short in-plane Mn-O bond distances and thus showing \textit{Q$_2$} as the dominant mode. To clearly identify \textit{Q$_2$} as the dominant mode or whether the \textit{Q$_2$} and \textit{Q$_3$} act cooperatively requires a method capable of capturing the insulating AFM and metallic FM behaviors within a single hybrid DFT functional.

\begin{table*}[]
\centering
\caption{$\Delta$E = E$_{AFM}$-E$_{FM}$ (meV), Mn magnetic moments ($\mu$ $_B$), energy gap (eV), and cell parameters (\r{A}) with their relative percentage errors in parenthesis when full structure relaxation has been performed. Our calculated PBE-10 and PBE-15 and the reported HSE-15 are the optimal hybrid functionals to reproduce the experimental properties at low T. }
\label{my-label2}
\begin{tabular}{lcccccc}
\\
\hline
         & $\Delta$E & m & a           & b           & c           & Gap      \\ \hline
PBE      & 172       & 3.62       & 5.540 (0.1) & 5.691 (-0.9) & 7.721 (-0.7) & metallic  \\
PBE-5   & 32.8       & 3.72      & 5.527 (-0.1) & 5.790 (0.8) & 7.672 (0.1) & 0.59    \\
PBE-10  & -51.4      & 3.76      & 5.521 (-0.2) & 5.810 (1.2) & 7.655 (-0.2) & 1.26    \\
PBE-15  & -53.2      & 3.80      & 5.514 (-0.3) & 5.832 (1.6) & 7.628 (-0.5) & 1.82    \\
PBE-20  & -66.9      & 3.82      & 5.509 (-0.4) & 5.825 (1.4) & 7.620 (-0.6) & 2.39    \\
PBE-25     & -50.0       & 3.85  & 5.501 (-0.6) & 5.810 (1.2) & 7.623 (-0.6) & 2.88    \\
B3LYP    & -18.1       & 3.80    & 5.561 (0.5)  & 5.937 (3.4) & 7.676 (0.1) & 2.61    \\ 
B3PW     & -64.1       & 3.82    & 5.514 (-0.3) & 5.844 (1.8) & 7.621 (-0.6) & 2.63    \\
HSE-25    & -23.6       & 3.85   & 5.501 (-0.6) & 5.811 (1.2) & 7.632 (-0.5) & 2.29    \\
Lc-$\omega$PBE & -109     & 3.71 & 5.550 (0.3)  & 5.897 (2.7) & 7.653 (-0.2) & 6.83    \\
HISS     & -57.4       & 3.83    & 5.492 (-0.7) & 5.761 (0.3) & 7.611 (-0.7) & 2.99    \\
M06      & -14.5       & 3.88    & 5.536 (0.1)  & 5.922 (3.1) & 7.643 (-0.3) & 2.58    \\ \hline  \hline
HSE-10$^a$  & 3        & 3.65    & 5.661 (2.3)  & 5.594 (-2.6) & 7.712 (0.6) & 0.75  \\
HSE-15$^a$  & -24       & 3.67   & 5.553 (0.4)  & 5.817 (1.3) & 7.633 (-0.5) & 1.63    \\
HSE-25$^a$     & -8       & 3.74 & 5.526 (-0.1) & 5.789 (0.8) & 7.628 (-0.5) & 2.47    \\
PBE+U(U=8,J=2)$^b$  & -2 & 3.67  & 5.507 (-0.6) & 5.822 (1.5) & 7.595 (-1.4) & 1.6    \\ \hline
Exp.      &     \textless 0        & 3.72$^{c}$,3.87$^{d}$  & 5.532$^{e}$ & 5.742$^{e}$  & 7.668$^{e}$       & 1.1$^{f}$,1.7$^{d}$,2.0$^{g}$ \\ \hline
\\ 
$^{a}$ \cite{franchini}
$^{b}$ \cite{mellan}
$^{c}$ \cite{exp1} \\
$^{d}$ \cite{exp2}
$^{e}$ \cite{exp0}
$^{f}$ \cite{exp3} \\
$^{g}$ \cite{exp4}

\end{tabular}
\end{table*}

The data obtained after full structural optimizations are shown in Tables II and III. Table II shows the AFM-FM ground state total energy differences relative to the FM state, the bandgaps, magnetic moments, and relaxed cell parameters. In Table III we show the long, short, and medium Mn-O bond lengths associated with the MnO$_6$ octahedra and the corresponding \textit{Q$_2$} and \textit{Q$_3$} vibrational modes. 


All hybrid functionals except PBE-5 and HSE-10 accurately describe the magnetic, electronic, and structural properties of the low T phase of LaMnO$_3$. These functionals yield a FM state while all other hybrid configurations lead to an insulating AFM state where the bandgaps increase with the percentage of HFX used. It should be noted that the long (Lc-$\omega$PBE) and triple (HISS) range separated functionals provide a large overestimation of the bandgaps. Therefore these functionals can be removed from consideration as possible candidates to capture the high temperature metallic phase. However, the bandgaps obtained by the traditional hybrid functionals (B3LYP and PBE0) and HSE, that uses 20$\%$ or more HFX mixing range between 2 and 3 eV, and are slightly overestimated. For this reason one should use a HFX mixing less than 20$\%$ to reproduce the bandgap and to improve the possibility of finding a single hybrid functional capable of accurately account for the low and high temperature phases. This is an important observation since, as previously mentioned, HFX favors an insulating behavior. In Fig. 3, it can be seen that mixing between 8$\%$ and 17$\%$ of HFX with PBE the calculated properties reproduce both the experimental bandgap and the AFM ordering. One can also see from Table II that the PBE-HFX functionals produce accurate lattice parameters compared to experimental data.

\begin{figure}[t]
\includegraphics[width=0.45\textwidth]{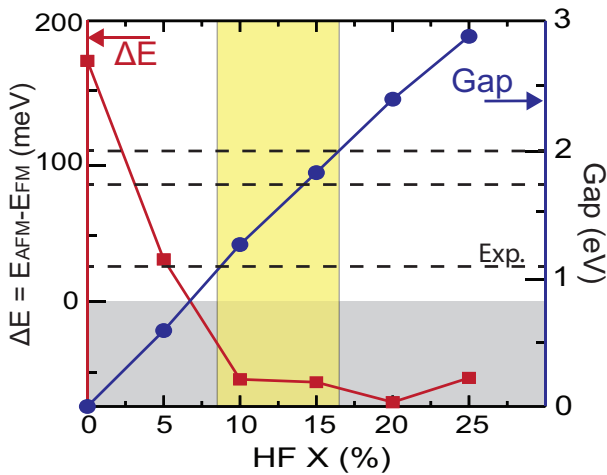}
\caption{$\Delta$=E$_{AFM}$-E$_{FM}$ and bandgap as a function of the HFX mixing. Solid lines connect the simulated results while dashed lines display the experimental bandgaps. This shows that only PBE-10 and PBE-15 can reproduce the magnetic ordering and bandgap of LaMnO$_3$.}
\label{fig3}
\end{figure}

\begin{figure}[t]
\includegraphics[width=0.40\textwidth]{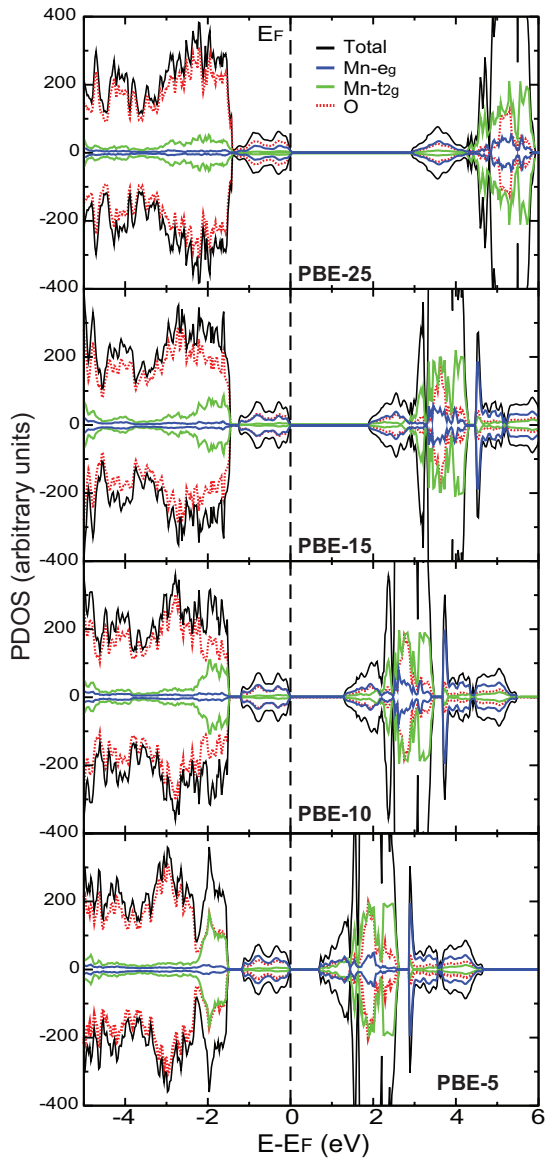}
\caption{(Color online) Projected density of states of fully relaxed AFM orthorhombic phase using PBE-5, PBE-10, PBE-15 and PBE-25 (PBE0) functionals. Only A-AFM phases are displayed. Solid black lines show the total density, blue and green are the 3d-e$_g$ and 3d-t$_{2g}$ levels respectively and red dashed lines are the O contributions. The bandgap is directly proportional to the percentage of HFX included in the functional. }
\label{fig4}
\end{figure}


Similar results were obtained by He and co-workers \cite{franchini,franchini2} when using a HFX mixing of less than 25$\%$ in the HSE functional. Both HSE-15 and PBE-10 produce an AFM insulating solution while HSE-10 and PBE-5 produce a FM metallic state. This behavior shows the importance of the HFX mixing contribution, where HFX in HSE-HFX decays exponentially at long-range while PBE-HFX contains the same HFX amount at any range. Therefore as the amount of HFX mixing decreases, HSE-HFX will produce the transition from insulator to metal before PBE-HFX. Including local HFX in the DFT exchange functional is important to reproduce the insulating state, but it should be noted that the decay of the exchange potential at long-range needs an appropriate balance between exponential screening and 1/r behavior to be able to accurately capture the metallic and insulating state properties (i.e., size of bandgap). This feature is also reflected in Table II where $a$ and $b$ lattice parameters for HSE-10 are significantly overestimated and underestimated respectively. In contrast, with HSE-15 these relative errors are highly reduced and the AFM insulator phase is then reproduced.

The AFM-FM energy differences are much smaller in HSE functionals than in PBE-HFX. In fact, HSE-15 produces the largest AFM-FM energy difference equal to -24 meV that corresponds to a temperature of $\simeq$ 279 K while for PBE-15 the energy difference corresponds to 615 K, which is much closer to the T$_{JT}$ = 750 K high temperature phase transition to the FM metallic state.  

The optimized cell parameters in Table II compare very well with the experimental values with relative errors inferior to 2$\%$ for all hybrid functionals except B3LYP, Lc-$\omega$PBE and M06. These three functionals produce slightly larger $b$ cell parameters, and an associated overestimation of the Mn-O distances. This directly affects the \textit{Q$_2$}, and \textit{Q$_3$} JT vibrational modes (see Table III). The origin of these errors may be attributed to the deficiency in the description of the exact homogeneous electron gas limit (B3LYP case) \cite{gaslimit} and the parametrization used in these functionals. For example, the empirical parameters contained in B3LYP or M06 were obtained by fitting properties of a set of systems that include a large number organic molecules but does not include oxide based materials.

\begin{table}[]
\centering
\caption{Calculated long (l), short (s), and medium (m) Mn-O bonds (\r{A}) (with their relative percentage errors in parenthesis) and \textit{Q$_2$} and \textit{Q$_3$} (\r{A}) Jahn Teller modes given by eq. 3 and 4.  }
\label{my-label3}
\begin{tabular}{lccccc}
\\
\hline
            & l     & s     & m     & \textit{Q$_2$}    & \textit{Q$_3$}     \\ \hline 
PBE-5      & 2.192 (0.4) & 1.926 (1.2) & 1.964 (0.4) & 0.376 & -0.155 \\
PBE-10     & 2.215 (1.4) & 1.917 (0.7) & 1.959 (0.1) & 0.421 & -0.175 \\
PBE-15     & 2.224 (1.8) & 1.912 (0.5) & 1.955 (-0.1) & 0.441 & -0.185 \\
PBE-20     & 2.224 (1.8) & 1.908 (0.3) & 1.953 (-0.2)& 0.447  & -0.185 \\
PBE-25     & 2.205 (1.0) & 1.902 (-0.1) & 1.971 (0.7) & 0.429 & -0.135 \\
B3LYP      & 2.293 (5.0) & 1.917 (0.7) & 1.971 (0.7) & 0.532 & -0.219  \\ 
B3PW       & 2.231 (2.2) & 1.907 (0.2) & 1.958 (0.1)& 0.458  & -0.181 \\
HSE-25     & 2.214 (1.4) & 1.907 (0.2) & 1.956 (0.1) & 0.434  & -0.171 \\
Lc-$\omega$PBE & 2.211 (1.2) & 1.880 (-1.2) & 1.943 (-0.7) & 0.468 & -0.167 \\
HISS        & 2.211 (1.2) & 1.888 (-0.8) & 1.900 (-0.3)  & 0.457 & -0.167 \\
M06         & 2.278 (4.3) & 1.901 (-0.1) & 2.000 (0.6) & 0.533 & -0.198 \\ \hline \hline
HSE-10$^a$  & 2.134 (-2.3) & 1.923 (1.1) & 1.979 (1.1) &  0.298 & -0.080    \\ 
HSE-15$^a$  & 2.213 (1.3)      & 1.914   (0.6)    & 1.962 (0.3) & 0.423 & -0.165  \\ 
HSE-25$^a$     & 2.214  (1.4)     & 1.905 (0.1)      & 1.957 (0.0) & 0.437 & -0.167    \\ \hline
Exp.$^b$         & 2.184 & 1.903 & 1.957 & 0.397 & -0.141 \\ \hline
\\
$^{a}$ \cite{franchini}, $^b$ \cite{exp0}

\end{tabular}
\end{table}

To demonstrate that our procedure yields accurate electronic states, we show the projected density of states (PDOS) for various PBE-HFX functionals in Fig. 4. The splitting observed between t$_{2g}$ and e$_{g}$ states is due to the crystal field splitting where the local symmetry of the MnO$_6$ octahedra promotes the e$_g$ states to lie higher in energy. The JT distortions produce another splitting that corresponds to a lifting of degeneracy of the 3d-e$_g$ level, leading to the opening of a bandgap. 

Comparing these results with the unrelaxed calculations in Table I, it can be seen that some functionals are quite sensitive to small variations of the structure. This leads to important changes in the magnetic ordering. We note that calculations using either PBE or PBE-5 lead to a change from AFM to FM upon relaxation while the opposite trend is found for M06. This demonstrates the sensitivity to small structural changes and the importance of performing full relaxation in order to obtain reliable electronic, magnetic and structural properties.

To summarize, from Fig. 3 and the results in the Tables I, II and III we are able to identify two hybrid functionals using 10$\%$ and 15$\%$ HFX mixing (PBE-10 and PBE-15) that accurately reproduce the experimental low temperature A-AFM phase, bandgap, and structural parameters.

\subsection{C. Study of the high T phase and the insulator-to-metal transition in LaMnO$_3$.}

Now that functional candidates have been identified for the low temperature phase, we examine if they are also able to properly describe the high temperature \textit{pseudo-cubic} phase. The structural change to a higher symmetry structure leads to nearly equal but slightly reduced $a$ and $b$ cell parameters with a slight increase in the $c$ parameter. In addition, $a$ $\sim$ $c$ /$\sqrt{2}$ and the system has a slight 0.36$\%$ decrease in volume \cite{Tapan}. To study the \textit{pseudo-cubic} high temperature phase of LaMnO$_3$, the crystal structure is approximated by setting the cell parameters as $a$ = $b$ = $c$ /$\sqrt{2}$.

We performed full geometry relaxations while constraining the cell parameters to effectively reduce the number of parameters to be optimized to one lattice parameter. We studied both FM and AFM (A-AFM, C-AFM and G-AFM) orderings to determine the lowest energy structure and magnetic ordering. The constraint applied does not affect other properties of the system including the rotations and tilts that alternate between MnO$_6$ octahedra as seen in Fig. 1, and lead to very small changes with relative errors smaller than 0.1$\%$ consistent with typical DFT errors in volume. In addition, the Pbnm and \textit{pseudo-cubic} structures are both subgroups of the cubic group and should have contributions to the free-energy due to phonons that are equal in magnitude. It follows that the internal energy (ground-state energy) should be the dominating term in the free-energy. Therefore, this approach provides a reliable model for studying the effect of temperature on the electronic structure using a ground-state simulation approach.

\begin{figure}[t]
\includegraphics[width=0.47 \textwidth]{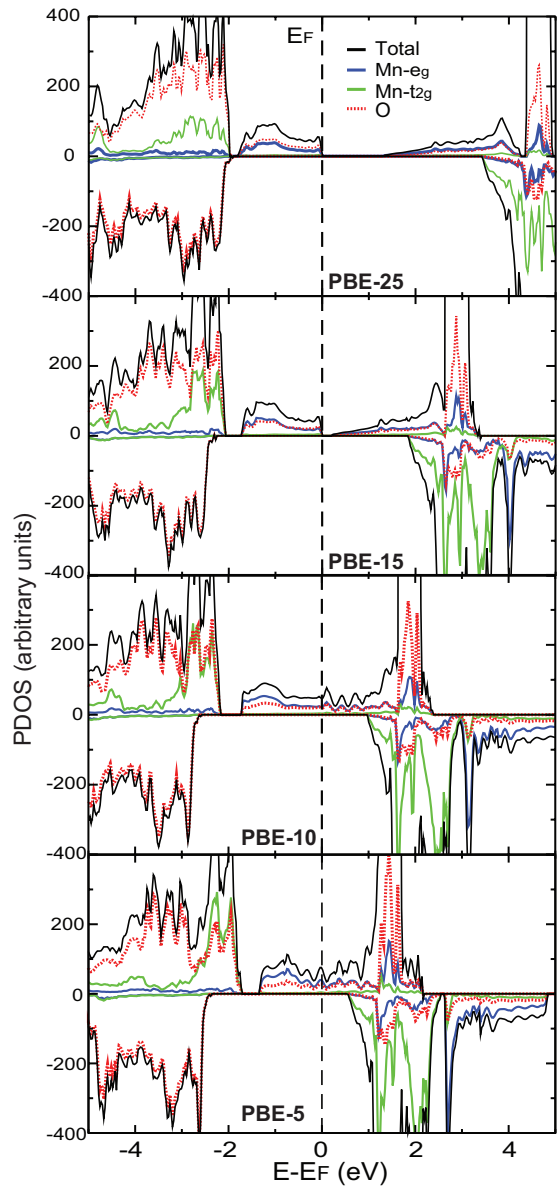}
\caption{(Color online) Projected density of states of the high T phase of LaMnO$_3$ via PBE-HFX hybrid functional. Only FM phases are displayed. PBE-5 and PBE-10 reproduce the experimental FM metallic phase of LaMnO$_3$ at high T. }
\label{fig6}
\end{figure}

\begin{figure}[t]
\includegraphics[width=0.48 \textwidth]{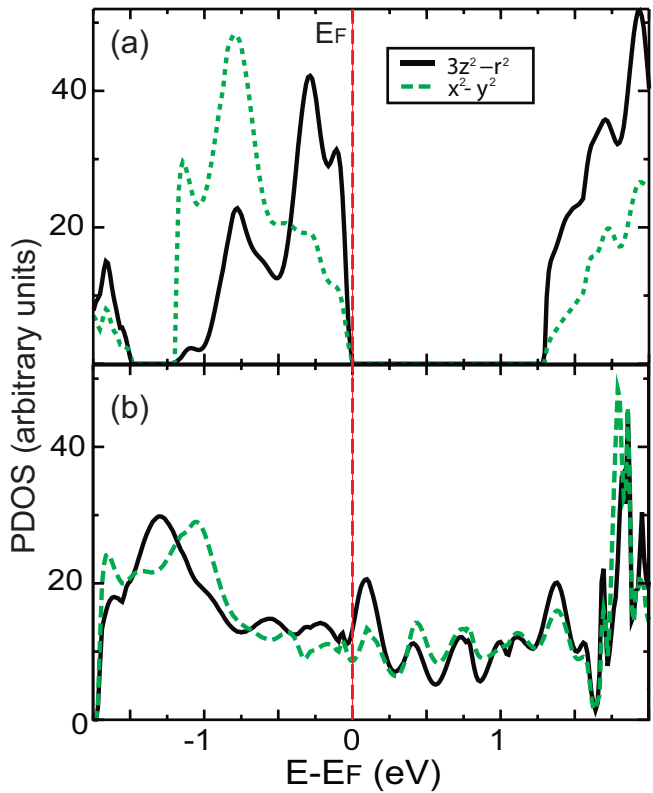}
\caption{ (Color online) Projected density of states ($\alpha$-$\beta$) of the Mn 3d-e$_g$ contributions in the vicinity of Fermi level for the low (a) and high (b) temperature phases. }
\label{fig8}
\end{figure}

\begin{table*}[]
\centering
\caption{Calculated cell parameters, Mn-O bond distances (\r{A}) (with their relative percentage errors in parenthesis), $\Delta$E = E$_{AFM}$-E$_{FM}$ (meV), bandgap (eV), and \textit{Q$_2$} and \textit{Q$_3$} JT vibrational modes (\r{A}) of the high T phase of LaMnO$_3$. PBE-10 and PBE-5 reproduce the experimental FM metallic state.}
\label{my-label4}
\begin{tabular}{lcccccccc}
\\
\hline
                      & Cell parameters     & l     & s     & m       & \textit{Q$_2$}    & \textit{Q$_3$}  & $\Delta$E       & Gap   \\ \hline
PBE-25       & 5.533 (0.9)              & 2.126 (-4.5) & 1.887 (5.1) & 1.999 (0.5)    & 0.338 & -0.012 & -6.7          & 2.45  \\
                                 
PBE-15         & 5.532   (0.9)            & 2.116 (-4.0) & 1.896 (4.6) & 1.998 (0.6)    & 0.311 & -0.013 & -19.1         & 0.14 \\
                              
PBE-10         & 5.547  (0.6)             & 2.077 (-2.1) & 1.941 (2.4) & 2.001 (0.4) & 0.192 & -0.013 & 40.0            & metallic\\
                              
PBE-5          & 5.544  (0.7)             & 2.020 (0.7) & 1.988 (0.0) & 1.996 (0.7) & 0.045 & -0.013 & 204         & metallic\\  \hline
          
Exp.$^a$                       & 5.582, 5.583, 7.890 & 2.034 & 1.988 & 2.010  & 0.066 & 0.001 &  \textgreater 0 & metallic \\ \hline
\\
$^a$ \cite{carvajal}

\end{tabular}
\end{table*}

Table IV shows the optimized \textit{pseudo-cubic} structural parameters, bandgaps and total energy differences along with the high temperature phase experimental results for comparison. It can be seen that PBE-10 and PBE-5 are in good agreement with experimental values. However, the use of higher values of HFX mixing yields an insulating solution.

In Fig. 5 the projected density of states for PBE-5 and PBE-10 exhibits a half-metallic character due to the strong Mn(3d-e$_g$)-O(p) hybridization that extends above the Fermi level. Note that this is the first time that a half-metallic character has been reported for the high temperature phase of LaMnO$_3$. The fact that slightly lower values of HFX provide the appropriate description for the electron delocalization in LaMnO$_3$ is not surprising since the functionals employed contain over 90$\%$ PBE exchange and provide a mainly exponential screening at long-range, which as previously mentioned, favors a metallic behavior. As the amount of HFX is increased the system should begin to transition to a more insulating type state as seen in Fig. 6. However, in the low temperature phase studies it was found that using a mixing of 5$\%$ of HFX or less leads to a metallic FM state. Thus, from the low and high temperature investigations an optimum value of 10$\%$ HFX mixing is chosen as the single hybrid functional that can accurately capture the electronic, magnetic and structural properties of both phases.

We note that all simulations performed in the \textit{pseudo-cubic} high T phase lead to a \textit{Q$_3$} JT mode that is nearly absent regardless of the electronic and magnetic state of the system (see Table IV). However, \textit{Q$_2$} mode shows an important reduction only when the FM metallic phase is reached. Therefore, our results indicate that \textit{Q$_2$} mode is the dominant mode responsible for driving the system through an IMT.

Fig. 6 shows the PDOS of the Mn 3d-e$_g$  states in the vicinity of the Fermi level for both low (a) and high (b) temperature phases. The JT distortions observed at low T are responsible for the additional energy splitting between e$_g$ levels. As the temperature rises above 750 K a reduction in the JT parameters drives the system into a \textit{pseudo-cubic} structure producing Mn sites with higher local symmetry as compared to the low temperature phase. This leads to a significant reduction in Mn-e$_g$ splitting, which produces nearly degenerate Mn-e$_g$ states. Therefore, temperatures above 750 K provide an extra degree of freedom for electrons where the probability of occupying either e$_g$ state is equal. This type of electron disorder can be viewed as a type of entropy of mixing similar to a high temperature homogeneously disordered 50-50 binary alloy where the probability of occupancy of either alloying species is the same. This effect lowers the free energy via an entropy of mixing type term (the orbital ordered phase has zero entropy). It should be note that octahedral tilts and rotations in the high T phase remain almost identical (with small decreases) compared to the low T phase as it can be seen in Fig. 1. At even higher temperatures the JT distortions disappear leading to regular MnO$_6$ octahedra with equal Mn-O bond distances \cite{qiu}.

To summarize, we have identified two hybrid functionals based on 5$\%$ and 10$\%$ HFX mixing with 95$\%$ and 90$\%$ of PBE exchange respectively (PBE-5 and PBE-10) that reproduce the high temperature FM metallic phase. However, only the PBE-10 hybrid functional can accurately account for both low and high T phases of LaMnO$_3$ where for the first time, we have uncovered a half-metallic FM system within a DFT framework.

\section{V. CONCLUSION}

We have shown that a single hybrid DFT functional with a particular HFX mixing can be used to describe systems where the electronic correlation depends on a range of conditions. This technique has been applied to LaMnO$_3$. LaMnO$_3$ is a particularly challenging material, since it contains strong coupling of spin, charge, and lattice degrees of freedom. This coupling is very sensitive to small changes in local structure where the distortions of the MnO$_6$ octahedra play an important role on the observed properties. By using 10$\%$ HFX with 90$\%$ PBE exchange we are able to accurately capture the effects of the JT distortions on the electronic, magnetic, and crystal structure on both low and high temperature phases.

Our simulations also capture the IMT observed in LaMnO$_3$ at high T and show that this transition has a strong dependence on the \textit{Q$_2$} JT mode due to its clear reduction on the MnO$_6$ octahedra. We also observe a FM half-metallic state consistent with previous investigations where hydrostatic pressure was applied, while in our case, we show that raising the temperature leads to a \textit{pseudo-cubic} structure with a reduced volume and a significant decrease of the JT distortions that produces a transition to the FM half-metallic state.

The computational work conducted by W.A.S. and P.R is supported by the U.S. Department of Energy under EPSCoR Grant No. DE-SC0012432 with additional support from the Louisiana Board of Regents and by an allocation of computing time from the Louisiana State University High Performance Computing center. V. M. acknowledges support by New York State under NYSTAR program C080117.

\end{document}